\documentclass[reprint, amsmath, amssymb, prl, cha, superscriptaddress]{revtex4-1}

\usepackage{graphicx}% Include figure files
\usepackage{dcolumn}% Align table columns on decimal point
\usepackage{bm}% bold math
\usepackage{gensymb}
\usepackage{pbox}

\begin{document}

\title{Minimization of atomic displacements as a guiding principle of the martensitic phase transformation}

\author{F\'elix Therrien}
 \affiliation{Colorado School of Mines}%Lines break automatically or can be forced with \\
\affiliation{National Renewable Energy Laboratory}
\author{Vladan Stevanovi\'c}%
 \email{vstevano@mines.edu}
\affiliation{Colorado School of Mines}%
\affiliation{National Renewable Energy Laboratory}

\date{\today}% It is always \today, today,
             %  but any date may be explicitly specified

\begin{abstract}
We present a unifying description for the martensitic transformation of steel that accounts for important experimentally observable features of the transformation namely, the Neumann bands, the interfacial (habit) plane between the transformed and untransformed phases and their orientation relationship (OR). It is obtained through a simple geometric minimization of the total distance traveled by all the atoms from the austenite (FCC or $\gamma$) phase to the martensite (BCC or $\alpha$) phase, without the need for any explicit energy minimization. Our description unites previously proposed mechanisms but it does not rely on assumptions and experimental knowledge regarding the shear planes and directions, or external adjustable parameters. We show how the Kurdjumov-Sach orientation relationship between the two phases and the $\{225\}_\gamma$ habit plane, which have both been extensively reported in experiments, naturally emerge from the distance minimization. We also propose an explanation for the occurrence of a different orientation relationship (Pitsch) in thin films.
\end{abstract}

\maketitle

Martensite, a body centered cubic (BCC) derived metastable phase of iron and carbon, often labeled $\alpha$, is typically obtained by a rapid quenching of the high temperature face-centered cubic (FCC) austenite phase ($\gamma$). The martensitic transformation ($\gamma \rightarrow \alpha$) is diffusionless and rapid. By virtue of its immense technological and industrial relevance, the mechanism behind the martensitic transformation has been extensively studied and theorized in the past.

Bain first proposed a mechanism for the transformation by establishing an atom-to-atom correspondence between the BCC and FCC lattices \cite{bain_1924}. Equivalent simple shear mechanisms were proposed subsequently \cite{kurdjumov1930mechanismus, nishiyama_2012} based on the measured orientation relationships (OR) between the martensite grains and the residual austenite, namely, the Kurdjumov-Sachs OR and Nishiyama-Wassermann OR. Other ORs have also been measured \cite{greninger1949mechanism, pitsch_1959}, in particular the Pitsch OR commonly observed in thin films \cite{olsen1971fcc, wuttig1993structural, kalki1993evidence}. These simple shear mechanisms failed to explain the measured interfacial planes (habit planes) between austenite and martensite as well as the observed $\{112\}_\alpha$ twinning \cite{mathewson1928, greninger1949mechanism, wayman1961crystallography, maki1977transmission, nishiyama_2012}. To account for these missing features the Phenomenological Theory of Martensitic Transformation (PTMT) \cite{greninger1949mechanism, jaswon1948atomic, bowles1951crystallographic, bowles1954crystallography, mackenzie1954crystallography, Wechsler1953, Wechsler1960} was developed. 
It assumes the existence of a plane that remains invariant during the transformation and is a consequence of a slipping or a twinning process. Although, the PTMT can explain observed habit planes and ORs, it relies on experimental information about the slipping/twinning process and ad-hoc adjustable paramters have to be introduced to reconcile observed shear and habit planes. Extensive descriptions of the PTMT and its history can be found elsewhere \cite{nishiyama_2012, khachaturyan2013theory, kelly2012crystallography}. Other theories were developped where the shape and orientation of domains of martensite in the austenite lattice can be determined by minimizing their elastic energy \cite{roitburd1969domain, roitburd1974theory, khachaturyan2013theory} or by minimizing the free energy assuming small strain (linear geometry) and a continuous interface (Hadarmard jump) \cite{ball1987, bhattacharya2003}. 
These theories can be used to study the evolution of the microstructure of various complex systems under different stress conditions and are the basis for more elaborate models \cite{levitas2009micromechanical1, levitas2009micromechanical2}. Yet, their crystallographic description of the transformation mechanism is equivalent to the PTMT. Others \cite{cayron_2013, cayron2015continuous, baur2017225, koumatos2017theoretical, koumatos2016optimality, koumatos2019parameter} have recently worked on alternative crystallographic models, but theoretical research on the subject has been mainly focused on molecular dynamics simulations, a detailed account of which can be found in Ref.~\cite{ou2017molecular}.

In our own work, we start from the assumption that the most likely transformation mechanism is the one that requires the least displacement of all the atoms in the crystal. This is a plausible physical assumption that was employed by Jaswon and Wheeler in their paper that lead to the PTMT \cite{jaswon1948atomic}, in our previous work on phase transformations \cite{stevanovic2018, therrien2019} and in the works of others. Let us consider the cumulative distance $d_1$ traveled by $N$ atoms from austenite to martensite:
\begin{equation}
d_1 = \sum_{l=1}^{m}\sum_{i,j,k = -\frac{n}{2}}^{\frac{n}{2}}||\mathbf{v}^{\alpha}_{ijkl} - \mathbf{v}^{\gamma}_{ijkl}|| \, , \label{eq:dist1}
\end{equation}
where $\mathbf{v}^{x}_{ijkl}$ are the position vectors of atoms of the initial austenite structure ($x=\alpha$) or the final martensite structure ($x=\gamma$) defined as 
$\mathbf{v}^{x}_{ijkl} = C^{x}(i, j, k) + \mathbf{p}_{l}^x$.
Here, $C^x$ is a unit cell of either the martensite or the austenite, and $\{\mathbf{p}_{l}^x:l=1 \dots m\}$ are the atomic positions inside $C^x$. The total number of atoms in this bloc of material is $N = m(n+1)^3$.
The choice of the unit cell vectors (including their orientation in space) and the order in which the atoms are indexed will determine the mechanism of the transformation and its corresponding $d_1$.
Thus, the actual mechanism of transformation, according to our initial assumption, can be characterized by the set of $C^x$ and $\{\mathbf{p}_{l}^x:l=1 \dots m\}$ that minimizes $d_1$ when $N \to \infty$. We have shown \cite{therrien2019} that the leading dependence of $d_1$ on the size ($N$) is a function of the distortion in the lattice, i.e., the strain. Therefore, by minimizing $d_1$, we also necessarily minimize the strain.

The Structure Matching Algorithm \cite{therrien2019} we developed recently and applied to this problem is an iterative approach to minimizing $d_1$. It directly minimizes the distance traveled by all ($N$) atoms belonging to a section of the crystal that is chosen to be as large as possible (ideally $N\rightarrow\infty$). After minimizing $d_1$ our algorithm retrieves the periodicity or the scale of the transformation ($C^x$) if it exists. We provide a brief description of the algorithm in the Supplementary Material (SM) and more details can be found in Ref.~\cite{therrien2019}. A full implementation is available on GitHub \footnote{github.com/ftherrien/p2ptrans}. We wish to emphasize the fact that the algorithm requires only the parameters of the initial and final structures as inputs. It relies entirely on the principle of minimal displacements, thus it does not directly take into account the energetics of the transformation. 
\begin{figure}
\includegraphics[width=\linewidth]{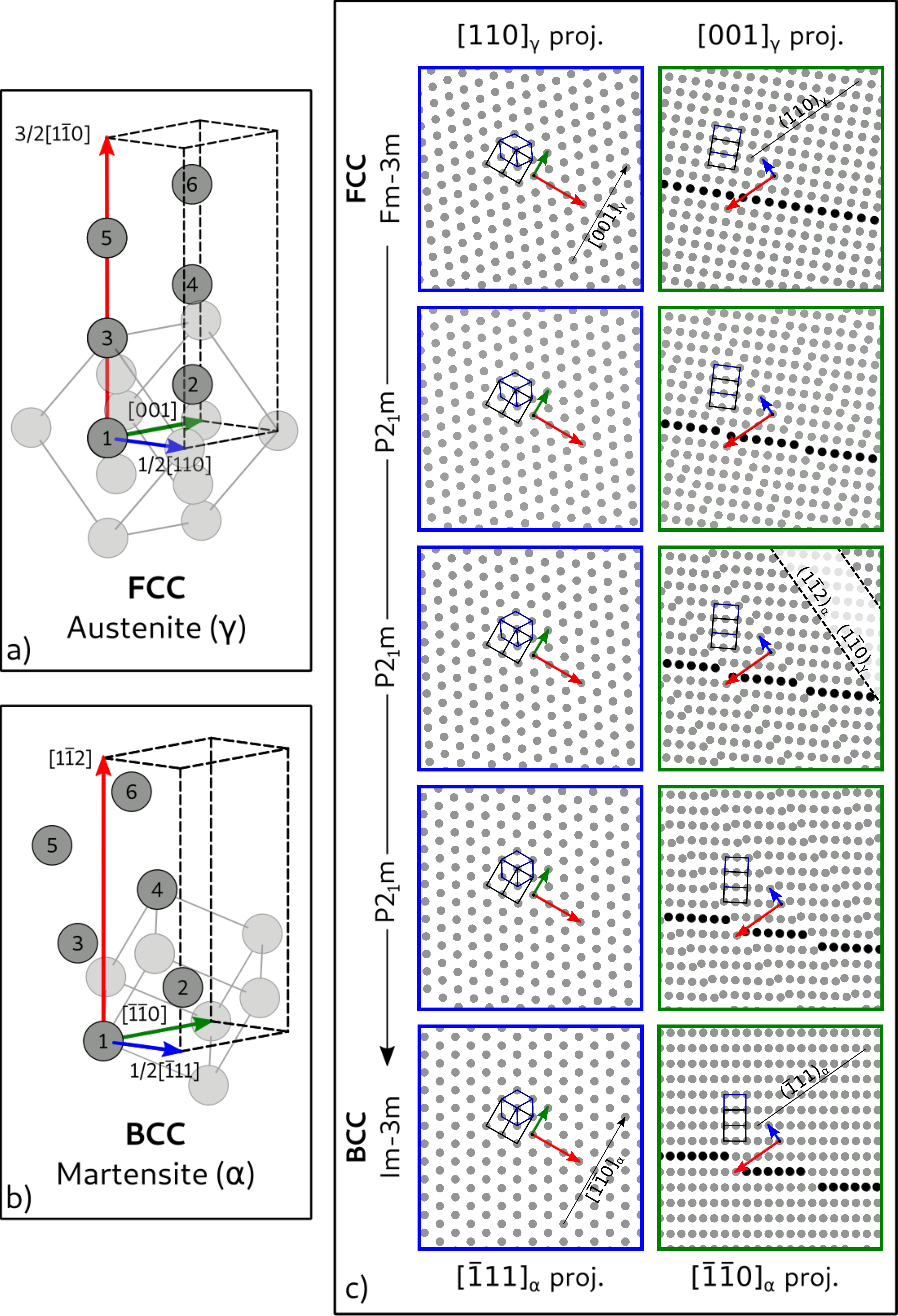}
\caption{\label{fig:martensitic}
Transformation cells of (a) austenite $C^\gamma$ and (b) martensite $C^\alpha$ are shown together with the atoms labeled according to the atom-to-atom correspondence between the two structures. The light gray atoms show the conventional cells. Panel (c) shows the evolution of the structure in 5 steps during the transformation from two different projections. In each image, the BCC conventional cell is represented in blue and the FCC conventional cell in black. In the $[\overline{1}\overline{1}0]_\alpha$ projection (green), one row of atoms is represented in black to help visualize the slipping process.}
\end{figure}

We used our distance minimization algorithm \cite{therrien2019} to find the transformation of pure iron from FCC to BCC. For austenite, we used the lattice constant $a_{\gamma}=3.57$\r{A}. For martensite we use both the hard-sphere packing lattice constant \cite{bogers1964partial,cayron_2013} ($a_{\alpha}=\sqrt{2/3} \, a_{\gamma}=2.915$\r{A}) and the experimentally measured lattice constant $a_{\alpha}=2.87$\r{A} and applied our algorithm to both choices. It is important to note that the lattice parameters themselves depend on the chemical element forming the crystal and that different elements have different lattice parameters which, in turn, lead to a different transformation mechanism.

Fig.~\ref{fig:martensitic} shows the transformation mechanism that minimizes $d_1$ resulting from our structure matching algorithm. The optimal $C^{\gamma}$ and $C^{\alpha}$ with their atomic positions are illustrated in panel (a) and (b) respectively.
From the $[110]_\gamma//[\overline{1}11]_\alpha$ perspective (blue frame in panel (c)) , going from FCC to BCC, one can see an elongation in the $[001]_\gamma//[\overline{1}\overline{1}0]_\alpha$ (green arrow) direction; it is particularly noticeable by looking at the change in shape of the FCC conventional cell in black. Now, looking in the $[001]_\gamma//[\overline{1}\overline{1}0]_\alpha$ direction (green frame), the transformation includes a shear of the $(1\overline{1}0)_\gamma//(1\overline{1}2)_\alpha$ plane in the $[110]_\gamma//[\overline{1}11]_\alpha$ direction with a slip every sixth layer.
The initial and final cells are linked by a transformation matrix $T$ such that $TC^{\gamma (\gamma)} = C^{\alpha (\gamma)}$ called the deformation gradient matrix (the exponent in parentheses indicates the basis). The matrix $T$ does not fully describe the transition because it does not account for the displacement of the atoms inside the cell. Indeed, from panel (a) to panel (b) in Fig.~\ref{fig:martensitic}, not only the cells have been distorted, but the atoms inside them have been displaced. This is why we find a mechanism of lower total atomic displacement then the Bain path despite the fact that it is optimal when considering only lattice deformation \cite{koumatos2016optimality}.
An animation of the transformation from the same viewing directions as in Fig.~\ref{fig:martensitic} and the evolution of the simulated X-ray diffraction patterns \cite{ong2013python} along the transformation are provided in SM together with the crystal structures (POSCAR format) for 60 snapshots along the transformation for both possible orientation relationships (explained further in the text).

Let us first analyze the deformation gradient matrix $T$. According to the polar decomposition theorem, it can always be written as $T=RU$ where $R$ is a rotation (unitary) matrix and $U$ is a symmetric matrix \cite{hall2015matrix}. Consequently $U-I$ is a proper strain tensor and its eigenvalues and eigenvectors are the principal strains and directions of the transformation. The eigenvalues of $U$ are the square roots of the eigenvalues of $T^TT$ and their eigenvectors are the same. Our structure matching algorithm \cite{therrien2019} gives us the optimal $T$ directly, from which one gets $U = P\,\text{diag}(\lambda_1,\lambda_2,\lambda_3)\,P^T$, where the columns of $P$ are the eigenvectors of $U$ in the basis of the FCC conventional cell.
Similarly, in the basis of the BCC lattice going from BCC to FCC, the eigenvalues are inverse and the eigenvectors form a different matrix Q.
The principal strains $\lambda_i - 1$ are: $2\sqrt{2}/3{-}1\approx-5.7\%$, $0\%$ and $\sqrt{4/3}{-}1\approx15.5\%$ using the hard-sphere packing lattice constant, and -7.2\%, 1.6\% and 13.7\% using the experimental lattice parameter. The strain directions, given by matrices $P$ and $Q$, are provided in SM. These strains are significantly lower than the one resulting from the Bain, Pitsch, Nishiyama-Wassermann (N-W) and Kurdjumov-Sachs (K-S) deformation paths which are $-18.4$\%, $15.5$\% and $15.5$\% (or -19.7\%, 13.7\%, 13.7\% using the experimental lattice parameter). The direction of the largest strain, $15.5$\%, in our solution is $[001]_\gamma//[\overline{1}\overline{1}0]_\alpha$ and therefore the two other principal strains lay in the plane perpendicular to it. In the Bain path, one of the two (degenerate) principal strains of 15.5\% can always be chosen to be in the same $[001]_\gamma//[\overline{1}\overline{1}0]_\alpha$ direction. Therefore, the difference between our proposed mechanism and the Bain distortion lies entirely in the $(001)_\gamma//(\overline{1}\overline{1}0)_\alpha$ plane. 

\begin{figure}
\includegraphics[width=\linewidth]{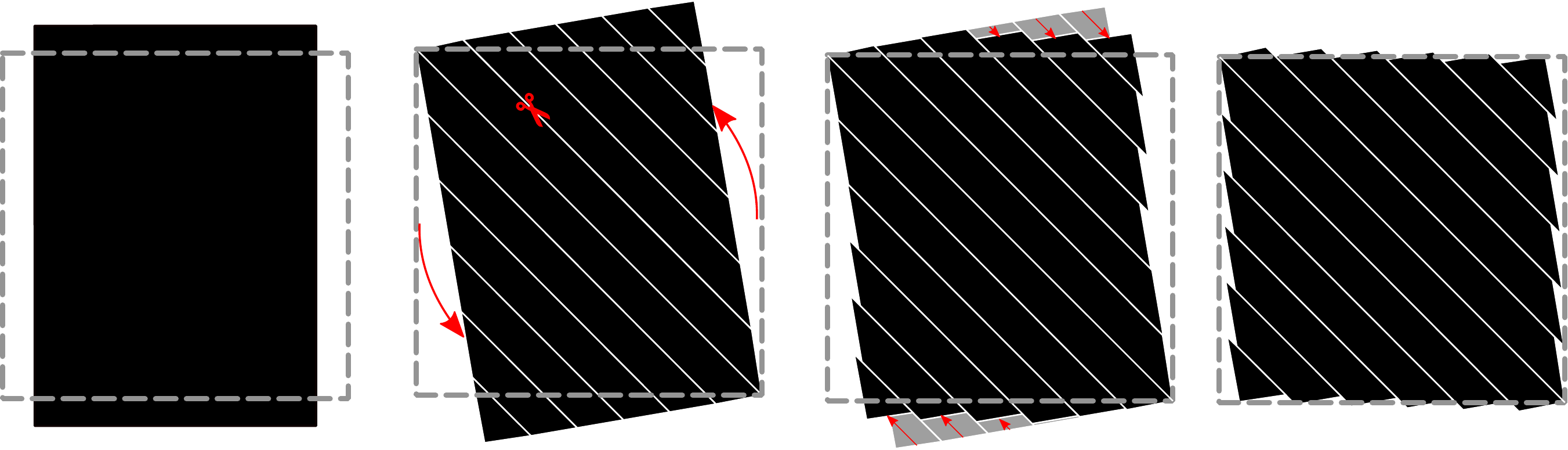}
\caption{\label{fig:macroslip}
Illustration of the effect of the slipping process. The dashed squares show the undistorted plane and the black rectangle on the left shows that same plane to which the Bain strains of -18,4\% and 15,5\% are applied. The steps to obtain a shape that is macroscopically similar to the undistorted plane but microscopically similar to the Bain-strained plane are shown from left to right.}
\end{figure}
The reduction of strain in that plane is due to slipping and naturally emerges from the $d_1$ minimization. Fig.~\ref{fig:macroslip} illustrates graphically how breaking down the plane in strips can reduce the macroscopic change in shape and therefore the strain. Imagine a sheet of metal that has been stretched from its original square shape by an amount corresponding to the Bain strains in the $(001)_\gamma//(\overline{1}\overline{1}0)_\alpha$ plane; its dimensions are now 1.155 by 0.816. By cutting the sheet in strips and by sliding them onto each other, one can obtain a shape that is much closer to the unstretched 1 by 1 sheet, while, locally, the metal in each strip has been stretched by an amount corresponding to the Bain strains (a similar argument for twinning is made in Fig. 2 of Ref.~\cite{roitburd1993}). In our proposed mechanism, the strips are 6 atomic layers wide and the actual transformation does not occur in two steps; each strip is distorted through a local shear that occurs simultaneously with the slipping process.

In order for this mechanism to yield a perfect BCC lattice, each strip needs to slip by an integer number of atomic layers as shown in Fig.~\ref{fig:martensitic}(c). Fulfilling this condition determines the width of the strips which is of 6 atomic layers using our choice of lattice parameters. In reality, martensite deviates from perfect BCC depending on the carbon content. We obtain qualitatively similar results if the tetragonal BCT martensite structure is used. It is important to note that this slipping process is fully described by the displacements of the atoms \textit{within} the transformation cells [Fig.~\ref{fig:martensitic} (a) and (b)]. This explains why our algorithm finds unit cell of 6 atoms; each atom  in the cell is displaced along the $[110]_\gamma$//$[\overline{1}11]_\alpha$ direction (blue cell vector) such that the condition is fulfilled at the end of the transformation.

The minimal distance result presented here bears striking resemblance to the PTMT since it involves a slipping process and, as discussed further, an invariant plane. 
In fact, in one of the original PTMT papers \cite{Wechsler1953} and in numerous subsequent studies \cite{wayman1961crystallography} the $\{112\}_\alpha$ plane is explicitly used as the slipping/twinning plane because striations parallel to that plane (sometimes referred to as Neumann bands) are commonly observed in martensite \cite{kelly1953neumann, maki1977transmission, shimizu1970association, nishiyama_2012, kelly2012crystallography}. The slipping process happens precisely along that $\{112\}_\alpha$ plane in our optimal distance mechanism.

Because we impose the final structure to be the perfect BCC lattice, our algorithm cannot find the related twinning process as it would lead to a different, twinned BCC lattice. However, by simply inverting the direction of the local displacements of the atoms for one column of unit cells along the $[110]_\gamma$//$[\overline{1}11]_\alpha$ direction as shown in Fig.~2 of the Supplementary Material (SM), we can obtain the twinned BCC. This would yield a mechanism very similar to the one presented in Ref.~\cite{baur2017225} but without the need to assume a particular OR. In that study, researchers also found the $\{11\sqrt{6}\}$ habit plane (discussed later) and found that the twinned and untwinned structures yield two variants of the Kurdjumov-Sach (K-S) orientation relationship.

Using the hard-sphere packing lattice constant, one of the principal strains is exactly zero, therefore, there necessarily exists a plane that is undistorted by the transformation.
However, using the experimental lattice parameter, there does not exist a plane that is fully invariant. Our approach is to look for a uniformly scaled plane instead of a fully invariant plane. Indeed, there always exists a plane such that the angles between its directions are preserved, i.e., a plane that is \textit{similar}, in the geometric sense, to the initial plane. Any vector $\mathbf{u}$ of that plane obeys
$||T\mathbf{u}|| = k||\mathbf{u}||$,
where k is a scalar, independent of the choice of $\mathbf{u}$. Ordering the eigenvalues of $U$ such that $\lambda_1 < \lambda_2 < \lambda_3$, one can show that the vectors $\mathbf{u}$ will form a plane if $k=\lambda_2$ (see SM). The lattice mismatch between the transformed plane and its equivalent in austenite is $1 - \lambda_2$. In the limit where $\lambda_2\to1$, which is the case when using the hard-sphere packing lattice constant, the mismatch is zero and the plane is invariant. Using $\lambda_{1,2,3}$ provided above and $P$, we find that the vector
${\mathbf{n}_{\text{HP}} = P(\sqrt{2}, 0, {\pm}\sqrt{6}) = (1, \overline{1}, {\pm}\sqrt{6}})$ is normal to the invariant plane.
This plane is approximately 0.5\degree~from the low index $(2\overline{2}{\pm}5)$ plane. Using the experimental parameters, we find the uniformly scaled plane to be about 0.4\degree~from the low index  $(2\overline{2}{\pm}5)$ plane with a mismatch of 1.6\%. This is an important result since the $\{225\}_\gamma$ habit plane is one of the few experimentally observed habit planes for low alloy plate-like martensite \cite{kelly2012crystallography, zhang_kelly_2005}. Moreover, interpreting the uniformly scaled plane as the habit plane allows us to readily obtain the $\{112\}_\alpha$ slipping process as well as the $\{225\}_\gamma$ habit plane without having to use a dilatation factor or additional shear processes which have been highly criticized by the detractors of the PTMT \cite{zhang_kelly_2005}.

Let us now consider that the uniformly scaled plane is also the habit plane between the two phases. In that case, the rotation $R$ is the one for which that plane does not rotate during the transformation. Thus, it must fulfill
${RU\mathbf{v}}/{||RU\mathbf{v}||} = \mathbf{v}$,
where $\mathbf{v}$ are unit vectors of the uniformly scaled plane. From that relation, using $U$, $P$, and $Q$ we can calculate the rotation matrix $R$. The exact steps necessary to obtain the matrices (one for $(1\overline{1}{+}\sqrt{6})$ and one for $(1\overline{1}{-}\sqrt{6})$)  
are detailed in SM.
The orientation relationship is given by the transformation matrix that changes the basis from FCC to BCC. Let us transform some vector $\mathbf{d}^{(\gamma)}$ from the FCC basis to the BCC basis: $R^T$ transforms the vector into the unrotated FCC basis, $P^T$ converts it to the eigenvalue basis and finally $Q$ converts it from the eigenvalue basis to the BCC basis. Hence:
$\mathbf{d}^{(\alpha)} = QP^TR^T\mathbf{d}^{(\gamma)}$.
By setting $\mathbf{d}^{(\gamma)}=[1\overline{1}1]_\gamma$ and using the first habit plane ($(1\overline{1}{+}\sqrt{6})$), we get $\mathbf{d}^{(\alpha)} = [0\,{-}\sqrt{3/2} \, \sqrt{3/2}]_\alpha$ which is parallel to $[0 \, \overline{1} \, 1]_\alpha$ and by setting $\mathbf{d}^{(\gamma)}=[110]_\gamma$ we get $\mathbf{d}^{(\alpha)} = [{-}\sqrt{2/3} \, \sqrt{2/3} \, \sqrt{2/3}]_\alpha$ which is parallel to $[\overline{1}11]_\alpha$. In other words, we find the following OR:
$[110](1\overline{1}1)_\gamma\text{//}[\overline{1}11](0\overline{1}1)_\alpha$
which is a variant of the K-S OR: the most commonly observed OR in plate-like martensite \cite{maki1977transmission}. Similarly, using the other habit plane ($(1\overline{1}{-}\sqrt{6})$), we get:
$[110](\overline{1}11)_\gamma\text{//}[\overline{1}11](\overline{1}0\overline{1})_\alpha$ 
which is another variant of the K-S OR. With the experimental lattice parameter, we find the same orientation relationship with a misalignment between the $\{111\}_\gamma$ and $\{011\}_\alpha$ planes of less than 0.4\degree.

Interestingly, if we assume that the mechanism is the same but that there is no extra rotation imposed by the habit plane ($R{=}0$ and $T{=}U$), the OR is given by:
$\mathbf{d}^{(\alpha)} = QP^T\mathbf{d}^{(\gamma)}$,
which leads to the $[001](110)_\gamma\text{//}[\overline{1}\overline{1}0](\overline{1}11)_\alpha$ orientation relationship; a variant of the Pitsch OR. This could explain why, in thin films, where the problem is reduced to two dimensions and 
the constraints imposed by the interfaces between austenite and martensite are less restrictive \cite{pitsch_1959, olsen1971fcc, bhattacharya2003}, it is the Pitsch OR (and not the K-S OR) that is observed experimentally \cite{pitsch_1959, olsen1971fcc, wuttig1993structural, kalki1993evidence}.

\begin{figure}
\includegraphics[width=\linewidth]{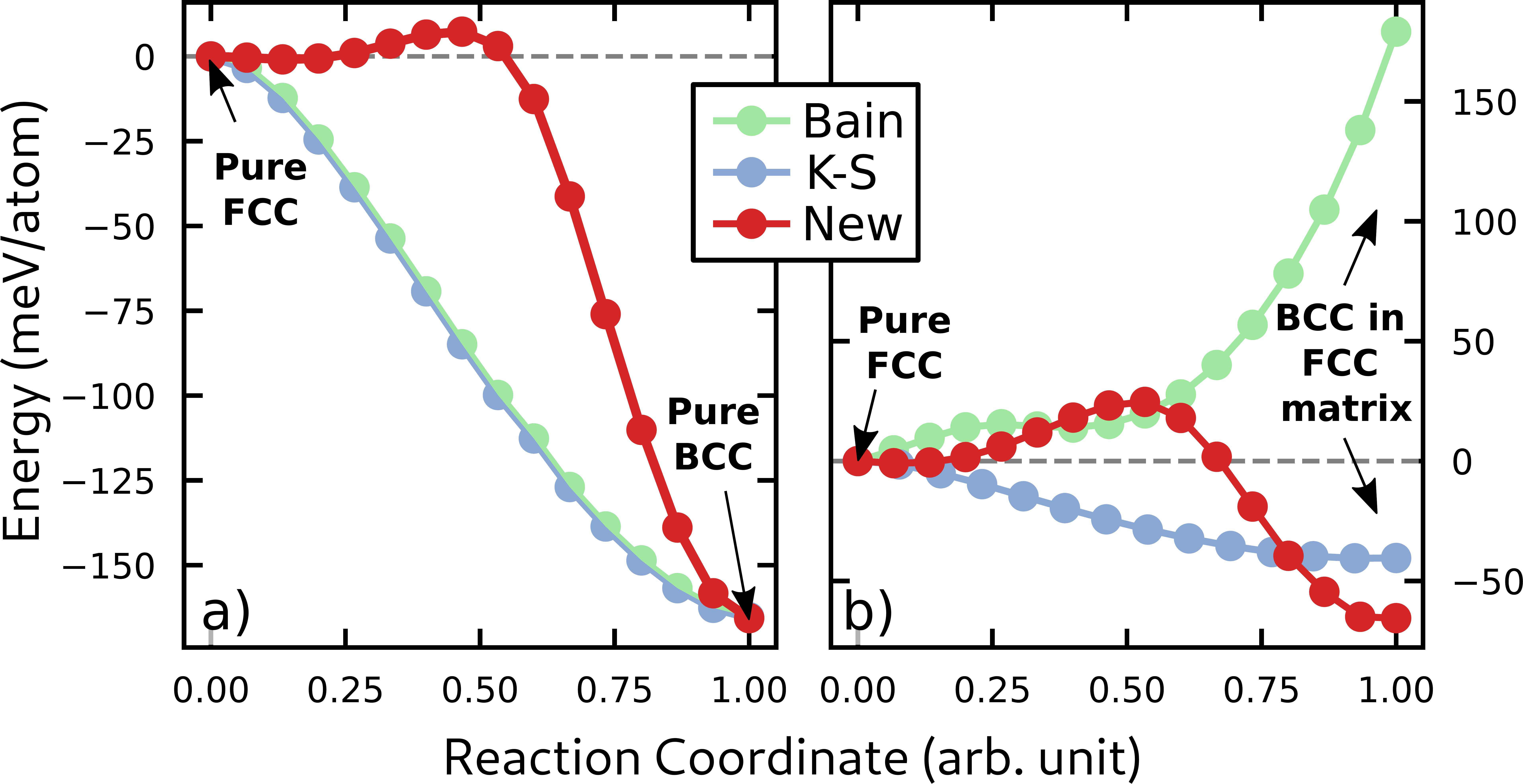}
\caption{\label{fig:energy}
Energy profile of the martensitic transformation. a) Transformation of an infinite iron single crystal from FCC (austenite) to BCC (martensite). b) Transformation of a plate of iron from FCC to BCC within the austenite matrix}
\end{figure}
In order to illustrate the consequences of our minimal displacements assumption, we computed energy profiles along the martensitic transformation from first principles (Fig.~\ref{fig:energy}). Without any phase coexistence (panel (a)), the Bain mechanism exhibits no barrier which makes it energetically advantageous over our proposed mechanism (New). However, in reality, austenite and martensite coexist both during and after the transformation. To account for that, we evaluated the energy profile of a thin (${\sim}2$~nm) infinite plate undergoing a martensitic transformation within a fixed austenite matrix. We compared: (1) a Bain mechanism in the Bain OR, (2) the Kurdjumov-Sach (K-S) shear mechanism (Bain path in the K-S OR) and (3) our proposed new mechanism in the K-S OR (New). Looking at Fig.~\ref{fig:energy}(b), it is clear that, despite an energy barrier caused by the slipping process, our optimal distance mechanism is the most energetically favorable in the final state due to its lower interfacial strain. The energy difference between the final states increases with plate thickness; hence, at a realistic thickness the final state emerging from the optimal distance mechanism would have significantly lower energy. The details of the spin-polarized density functional calculations and the corresponding crystal structures are given in the Supplementary Materials (SM) which includes Refs.~\cite{blochl1994projector, kresse1996efficiency}.

In conclusion, we showed how minimizing the distance \cite{therrien2019} traveled by all atoms from the austenite to the martensite phase in steels provides a description of the martensitic transformation. It can explain the key experimentally observable features of the transformation without relying on any experimental input (except lattice constants) and without any adjustable parameters. Our description is unifying in the sense that, using nothing but the principle of minimal displacements of the atoms, it naturally incorporates several elements of previous theories including the assumption initially made in Ref.~\cite{jaswon1948atomic}, the slipping and twinning processes found in Refs.~\cite{Wechsler1953, bowles1951crystallographic, ball1987, roitburd1974theory} and subsequent work, the habit plane and mechanism found in Ref.~\cite{baur2017225} and the fundamental role of the Pitsch mechanism described in Refs.~\cite{pitsch_1959, cayron_2013}. We thereby presented a simple solution to the long-standing and important problem of finding a general description of the martensitic transformation.
Our results suggest that distance minimization on its own can be relevant to describe certain diffusionless, solid-solid phase transformations. Hence, since our structure matching procedure is not specific to the martensitic transformation, it could potentially be used to study more complex systems such as shape memory alloys or solid-solid phase change materials. Moreover, there is a promising outlook for the use of our methodology in the field of interface physics as many models based on geometric principles have already been successful in describing interfaces \cite{bollmann1, balluffi_brokman_king_1982, zur1984, ikuhara_pirouz_1996, tkatchenko2004unequal, tkatchenko2005unequal, tkatchenko2006classification, persson_ACS:2016, jelver2017determination}.

\begin{acknowledgments}
This work is supported by the National Science Foundation, grant No. DMR-1945010 and was performed using computational resources sponsored by the Department of Energy's Office of Energy Efficiency and Renewable Energy, located at the National Renewable Energy Laboratory.
\end{acknowledgments}

%merlin.mbs apsrev4-1.bst 2010-07-25 4.21a (PWD, AO, DPC) hacked
%Control: key (0)
%Control: author (8) initials jnrlst
%Control: editor formatted (1) identically to author
%Control: production of article title (-1) disabled
%Control: page (0) single
%Control: year (1) truncated
%Control: production of eprint (0) enabled
%

%\bibliography{reference}% Produces the bibliography via BibTeX.

\begin{thebibliography}{53}%
\makeatletter
\providecommand \@ifxundefined [1]{%
 \@ifx{#1\undefined}
}%
\providecommand \@ifnum [1]{%
 \ifnum #1\expandafter \@firstoftwo
 \else \expandafter \@secondoftwo
 \fi
}%
\providecommand \@ifx [1]{%
 \ifx #1\expandafter \@firstoftwo
 \else \expandafter \@secondoftwo
 \fi
}%
\providecommand \natexlab [1]{#1}%
\providecommand \enquote  [1]{``#1''}%
\providecommand \bibnamefont  [1]{#1}%
\providecommand \bibfnamefont [1]{#1}%
\providecommand \citenamefont [1]{#1}%
\providecommand \href@noop [0]{\@secondoftwo}%
\providecommand \href [0]{\begingroup \@sanitize@url \@href}%
\providecommand \@href[1]{\@@startlink{#1}\@@href}%
\providecommand \@@href[1]{\endgroup#1\@@endlink}%
\providecommand \@sanitize@url [0]{\catcode `\\12\catcode `\$12\catcode
  `\&12\catcode `\#12\catcode `\^12\catcode `\_12\catcode `\%12\relax}%
\providecommand \@@startlink[1]{}%
\providecommand \@@endlink[0]{}%
\providecommand \url  [0]{\begingroup\@sanitize@url \@url }%
\providecommand \@url [1]{\endgroup\@href {#1}{\urlprefix }}%
\providecommand \urlprefix  [0]{URL }%
\providecommand \Eprint [0]{\href }%
\providecommand \doibase [0]{http://dx.doi.org/}%
\providecommand \selectlanguage [0]{\@gobble}%
\providecommand \bibinfo  [0]{\@secondoftwo}%
\providecommand \bibfield  [0]{\@secondoftwo}%
\providecommand \translation [1]{[#1]}%
\providecommand \BibitemOpen [0]{}%
\providecommand \bibitemStop [0]{}%
\providecommand \bibitemNoStop [0]{.\EOS\space}%
\providecommand \EOS [0]{\spacefactor3000\relax}%
\providecommand \BibitemShut  [1]{\csname bibitem#1\endcsname}%
\let\auto@bib@innerbib\@empty
%</preamble>
\bibitem [{\citenamefont {Bain}\ and\ \citenamefont
  {Dunkirk}(1924)}]{bain_1924}%
  \BibitemOpen
  \bibfield  {author} {\bibinfo {author} {\bibfnamefont {E.~C.}\ \bibnamefont
  {Bain}}\ and\ \bibinfo {author} {\bibfnamefont {N.}~\bibnamefont {Dunkirk}},\
  }\href@noop {} {\bibfield  {journal} {\bibinfo  {journal} {trans. AIME}\
  }\textbf {\bibinfo {volume} {70}},\ \bibinfo {pages} {25} (\bibinfo {year}
  {1924})}\BibitemShut {NoStop}%
\bibitem [{\citenamefont {Kurdjumov}\ and\ \citenamefont
  {Sachs}(1930)}]{kurdjumov1930mechanismus}%
  \BibitemOpen
  \bibfield  {author} {\bibinfo {author} {\bibfnamefont {G.}~\bibnamefont
  {Kurdjumov}}\ and\ \bibinfo {author} {\bibfnamefont {G.}~\bibnamefont
  {Sachs}},\ }\href@noop {} {\bibfield  {journal} {\bibinfo  {journal}
  {Zeitschrift f{\"u}r Physik}\ }\textbf {\bibinfo {volume} {64}},\ \bibinfo
  {pages} {325} (\bibinfo {year} {1930})}\BibitemShut {NoStop}%
\bibitem [{\citenamefont {Nishiyama}(2012)}]{nishiyama_2012}%
  \BibitemOpen
  \bibfield  {author} {\bibinfo {author} {\bibfnamefont {Z.}~\bibnamefont
  {Nishiyama}},\ }\href@noop {} {\emph {\bibinfo {title} {Martensitic
  transformation}}}\ (\bibinfo  {publisher} {Elsevier},\ \bibinfo {year}
  {2012})\BibitemShut {NoStop}%
\bibitem [{\citenamefont {Greninger}\ and\ \citenamefont
  {Troiano}(1949)}]{greninger1949mechanism}%
  \BibitemOpen
  \bibfield  {author} {\bibinfo {author} {\bibfnamefont {A.~B.}\ \bibnamefont
  {Greninger}}\ and\ \bibinfo {author} {\bibfnamefont {A.~R.}\ \bibnamefont
  {Troiano}},\ }\href@noop {} {\bibfield  {journal} {\bibinfo  {journal} {JOM}\
  }\textbf {\bibinfo {volume} {1}},\ \bibinfo {pages} {590} (\bibinfo {year}
  {1949})}\BibitemShut {NoStop}%
\bibitem [{\citenamefont {Pitsch}(1959)}]{pitsch_1959}%
  \BibitemOpen
  \bibfield  {author} {\bibinfo {author} {\bibfnamefont {W.}~\bibnamefont
  {Pitsch}},\ }\href@noop {} {\bibfield  {journal} {\bibinfo  {journal}
  {Philosophical Magazine}\ }\textbf {\bibinfo {volume} {4}},\ \bibinfo {pages}
  {577} (\bibinfo {year} {1959})}\BibitemShut {NoStop}%
\bibitem [{\citenamefont {Olsen}\ and\ \citenamefont
  {Jesser}(1971)}]{olsen1971fcc}%
  \BibitemOpen
  \bibfield  {author} {\bibinfo {author} {\bibfnamefont {G.}~\bibnamefont
  {Olsen}}\ and\ \bibinfo {author} {\bibfnamefont {W.}~\bibnamefont {Jesser}},\
  }\href@noop {} {\bibfield  {journal} {\bibinfo  {journal} {Acta
  Metallurgica}\ }\textbf {\bibinfo {volume} {19}},\ \bibinfo {pages} {1009}
  (\bibinfo {year} {1971})}\BibitemShut {NoStop}%
\bibitem [{\citenamefont {Wuttig}\ \emph {et~al.}(1993)\citenamefont {Wuttig},
  \citenamefont {Feldmann}, \citenamefont {Thomassen}, \citenamefont {May},
  \citenamefont {Zillgen}, \citenamefont {Brodde}, \citenamefont {Hannemann},\
  and\ \citenamefont {Neddermeyer}}]{wuttig1993structural}%
  \BibitemOpen
  \bibfield  {author} {\bibinfo {author} {\bibfnamefont {M.}~\bibnamefont
  {Wuttig}}, \bibinfo {author} {\bibfnamefont {B.}~\bibnamefont {Feldmann}},
  \bibinfo {author} {\bibfnamefont {J.}~\bibnamefont {Thomassen}}, \bibinfo
  {author} {\bibfnamefont {F.}~\bibnamefont {May}}, \bibinfo {author}
  {\bibfnamefont {H.}~\bibnamefont {Zillgen}}, \bibinfo {author} {\bibfnamefont
  {A.}~\bibnamefont {Brodde}}, \bibinfo {author} {\bibfnamefont
  {H.}~\bibnamefont {Hannemann}}, \ and\ \bibinfo {author} {\bibfnamefont
  {H.}~\bibnamefont {Neddermeyer}},\ }\href@noop {} {\bibfield  {journal}
  {\bibinfo  {journal} {Surface science}\ }\textbf {\bibinfo {volume} {291}},\
  \bibinfo {pages} {14} (\bibinfo {year} {1993})}\BibitemShut {NoStop}%
\bibitem [{\citenamefont {Kalki}\ \emph {et~al.}(1993)\citenamefont {Kalki},
  \citenamefont {Chambliss}, \citenamefont {Johnson}, \citenamefont {Wilson},\
  and\ \citenamefont {Chiang}}]{kalki1993evidence}%
  \BibitemOpen
  \bibfield  {author} {\bibinfo {author} {\bibfnamefont {K.}~\bibnamefont
  {Kalki}}, \bibinfo {author} {\bibfnamefont {D.}~\bibnamefont {Chambliss}},
  \bibinfo {author} {\bibfnamefont {K.}~\bibnamefont {Johnson}}, \bibinfo
  {author} {\bibfnamefont {R.}~\bibnamefont {Wilson}}, \ and\ \bibinfo {author}
  {\bibfnamefont {S.}~\bibnamefont {Chiang}},\ }\href@noop {} {\bibfield
  {journal} {\bibinfo  {journal} {Physical Review B}\ }\textbf {\bibinfo
  {volume} {48}},\ \bibinfo {pages} {18344} (\bibinfo {year}
  {1993})}\BibitemShut {NoStop}%
\bibitem [{\citenamefont {Mathewson}\ and\ \citenamefont
  {Edmunds}(1928)}]{mathewson1928}%
  \BibitemOpen
  \bibfield  {author} {\bibinfo {author} {\bibfnamefont {C.~H.}\ \bibnamefont
  {Mathewson}}\ and\ \bibinfo {author} {\bibfnamefont {G.}~\bibnamefont
  {Edmunds}},\ }\href@noop {} {\bibfield  {journal} {\bibinfo  {journal}
  {Transactions of the American Institue of Mining and Metallurgical
  Engineers}\ }\textbf {\bibinfo {volume} {80}},\ \bibinfo {pages} {311}
  (\bibinfo {year} {1928})}\BibitemShut {NoStop}%
\bibitem [{\citenamefont {Wayman}\ \emph {et~al.}(1961)\citenamefont {Wayman},
  \citenamefont {Hanafee},\ and\ \citenamefont
  {Read}}]{wayman1961crystallography}%
  \BibitemOpen
  \bibfield  {author} {\bibinfo {author} {\bibfnamefont {C.}~\bibnamefont
  {Wayman}}, \bibinfo {author} {\bibfnamefont {J.}~\bibnamefont {Hanafee}}, \
  and\ \bibinfo {author} {\bibfnamefont {T.}~\bibnamefont {Read}},\ }\href@noop
  {} {\bibfield  {journal} {\bibinfo  {journal} {Acta Metallurgica}\ }\textbf
  {\bibinfo {volume} {9}},\ \bibinfo {pages} {391} (\bibinfo {year}
  {1961})}\BibitemShut {NoStop}%
\bibitem [{\citenamefont {Maki}\ and\ \citenamefont
  {Wayman}(1977)}]{maki1977transmission}%
  \BibitemOpen
  \bibfield  {author} {\bibinfo {author} {\bibfnamefont {T.}~\bibnamefont
  {Maki}}\ and\ \bibinfo {author} {\bibfnamefont {C.}~\bibnamefont {Wayman}},\
  }\href@noop {} {\bibfield  {journal} {\bibinfo  {journal} {Acta
  Metallurgica}\ }\textbf {\bibinfo {volume} {25}},\ \bibinfo {pages} {681}
  (\bibinfo {year} {1977})}\BibitemShut {NoStop}%
\bibitem [{\citenamefont {Jaswon}\ and\ \citenamefont
  {Wheeler}(1948)}]{jaswon1948atomic}%
  \BibitemOpen
  \bibfield  {author} {\bibinfo {author} {\bibfnamefont {M.}~\bibnamefont
  {Jaswon}}\ and\ \bibinfo {author} {\bibfnamefont {J.}~\bibnamefont
  {Wheeler}},\ }\href@noop {} {\bibfield  {journal} {\bibinfo  {journal} {Acta
  Crystallographica}\ }\textbf {\bibinfo {volume} {1}},\ \bibinfo {pages} {216}
  (\bibinfo {year} {1948})}\BibitemShut {NoStop}%
\bibitem [{\citenamefont {Bowles}(1951)}]{bowles1951crystallographic}%
  \BibitemOpen
  \bibfield  {author} {\bibinfo {author} {\bibfnamefont {J.}~\bibnamefont
  {Bowles}},\ }\href@noop {} {\bibfield  {journal} {\bibinfo  {journal} {Acta
  Crystallographica}\ }\textbf {\bibinfo {volume} {4}},\ \bibinfo {pages} {162}
  (\bibinfo {year} {1951})}\BibitemShut {NoStop}%
\bibitem [{\citenamefont {Bowles}\ and\ \citenamefont
  {Mackenzie}(1954)}]{bowles1954crystallography}%
  \BibitemOpen
  \bibfield  {author} {\bibinfo {author} {\bibfnamefont {J.}~\bibnamefont
  {Bowles}}\ and\ \bibinfo {author} {\bibfnamefont {J.}~\bibnamefont
  {Mackenzie}},\ }\href@noop {} {\bibfield  {journal} {\bibinfo  {journal}
  {Acta metallurgica}\ }\textbf {\bibinfo {volume} {2}},\ \bibinfo {pages}
  {129} (\bibinfo {year} {1954})}\BibitemShut {NoStop}%
\bibitem [{\citenamefont {Mackenzie}\ and\ \citenamefont
  {Bowles}(1954)}]{mackenzie1954crystallography}%
  \BibitemOpen
  \bibfield  {author} {\bibinfo {author} {\bibfnamefont {J.}~\bibnamefont
  {Mackenzie}}\ and\ \bibinfo {author} {\bibfnamefont {J.}~\bibnamefont
  {Bowles}},\ }\href@noop {} {\bibfield  {journal} {\bibinfo  {journal} {Acta
  Metallurgica}\ }\textbf {\bibinfo {volume} {2}},\ \bibinfo {pages} {138}
  (\bibinfo {year} {1954})}\BibitemShut {NoStop}%
\bibitem [{\citenamefont {Wechsler}\ \emph {et~al.}(1953)\citenamefont
  {Wechsler}, \citenamefont {Lieberman},\ and\ \citenamefont
  {TA}}]{Wechsler1953}%
  \BibitemOpen
  \bibfield  {author} {\bibinfo {author} {\bibfnamefont {M.}~\bibnamefont
  {Wechsler}}, \bibinfo {author} {\bibfnamefont {D.}~\bibnamefont {Lieberman}},
  \ and\ \bibinfo {author} {\bibfnamefont {R.}~\bibnamefont {TA}},\ }\href@noop
  {} {\bibfield  {journal} {\bibinfo  {journal} {Transactions of the American
  Institue of Mining and Metallurgical Engineers}\ }\textbf {\bibinfo {volume}
  {197}},\ \bibinfo {pages} {1503} (\bibinfo {year} {1953})}\BibitemShut
  {NoStop}%
\bibitem [{\citenamefont {Wechsler}\ \emph {et~al.}(1960)\citenamefont
  {Wechsler}, \citenamefont {TA},\ and\ \citenamefont
  {Lieberman}}]{Wechsler1960}%
  \BibitemOpen
  \bibfield  {author} {\bibinfo {author} {\bibfnamefont {M.}~\bibnamefont
  {Wechsler}}, \bibinfo {author} {\bibfnamefont {R.}~\bibnamefont {TA}}, \ and\
  \bibinfo {author} {\bibfnamefont {D.}~\bibnamefont {Lieberman}},\ }\href@noop
  {} {\bibfield  {journal} {\bibinfo  {journal} {Transactions of the American
  Institue of Mining and Metallurgical Engineers}\ }\textbf {\bibinfo {volume}
  {218}},\ \bibinfo {pages} {202} (\bibinfo {year} {1960})}\BibitemShut
  {NoStop}%
\bibitem [{\citenamefont {Khachaturyan}(2013)}]{khachaturyan2013theory}%
  \BibitemOpen
  \bibfield  {author} {\bibinfo {author} {\bibfnamefont {A.~G.}\ \bibnamefont
  {Khachaturyan}},\ }\href@noop {} {\emph {\bibinfo {title} {Theory of
  structural transformations in solids}}}\ (\bibinfo  {publisher} {Courier
  Corporation},\ \bibinfo {year} {2013})\BibitemShut {NoStop}%
\bibitem [{\citenamefont {Kelly}(2012)}]{kelly2012crystallography}%
  \BibitemOpen
  \bibfield  {author} {\bibinfo {author} {\bibfnamefont {P.}~\bibnamefont
  {Kelly}},\ }in\ \href@noop {} {\emph {\bibinfo {booktitle} {Phase
  transformations in steels}}}\ (\bibinfo  {publisher} {Elsevier},\ \bibinfo
  {year} {2012})\ pp.\ \bibinfo {pages} {3--33}\BibitemShut {NoStop}%
\bibitem [{\citenamefont {Ro{\u\i}tburd}(1968)}]{roitburd1969domain}%
  \BibitemOpen
  \bibfield  {author} {\bibinfo {author} {\bibfnamefont {A.}~\bibnamefont
  {Ro{\u\i}tburd}},\ }\href@noop {} {\bibfield  {journal} {\bibinfo  {journal}
  {Soviet Physics Solid State}\ }\textbf {\bibinfo {volume} {10}},\ \bibinfo
  {pages} {2870} (\bibinfo {year} {1968})}\BibitemShut {NoStop}%
\bibitem [{\citenamefont {Ro{\u\i}tburd}(1974)}]{roitburd1974theory}%
  \BibitemOpen
  \bibfield  {author} {\bibinfo {author} {\bibfnamefont {A.}~\bibnamefont
  {Ro{\u\i}tburd}},\ }\href@noop {} {\bibfield  {journal} {\bibinfo  {journal}
  {Soviet Physics Uspekhi}\ }\textbf {\bibinfo {volume} {17}},\ \bibinfo
  {pages} {326} (\bibinfo {year} {1974})}\BibitemShut {NoStop}%
\bibitem [{\citenamefont {Ball}\ and\ \citenamefont {James}(1987)}]{ball1987}%
  \BibitemOpen
  \bibfield  {author} {\bibinfo {author} {\bibfnamefont {J.}~\bibnamefont
  {Ball}}\ and\ \bibinfo {author} {\bibfnamefont {R.}~\bibnamefont {James}},\
  }\href@noop {} {\bibfield  {journal} {\bibinfo  {journal} {Archive for
  Rational Mechanics and Analysis}\ }\textbf {\bibinfo {volume} {100}},\
  \bibinfo {pages} {13} (\bibinfo {year} {1987})}\BibitemShut {NoStop}%
\bibitem [{\citenamefont {Bhattacharya}(2003)}]{bhattacharya2003}%
  \BibitemOpen
  \bibfield  {author} {\bibinfo {author} {\bibfnamefont {K.}~\bibnamefont
  {Bhattacharya}},\ }\href@noop {} {\emph {\bibinfo {title} {Microstructure of
  martensite : why it forms and how it gives rise to the shape-memory
  effect}}},\ Oxford series on materials modelling ; 2\ (\bibinfo  {publisher}
  {Oxford University Press},\ \bibinfo {address} {Oxford ;},\ \bibinfo {year}
  {2003})\BibitemShut {NoStop}%
\bibitem [{\citenamefont {Levitas}\ and\ \citenamefont
  {Ozsoy}(2009{\natexlab{a}})}]{levitas2009micromechanical1}%
  \BibitemOpen
  \bibfield  {author} {\bibinfo {author} {\bibfnamefont {V.~I.}\ \bibnamefont
  {Levitas}}\ and\ \bibinfo {author} {\bibfnamefont {I.~B.}\ \bibnamefont
  {Ozsoy}},\ }\href@noop {} {\bibfield  {journal} {\bibinfo  {journal}
  {International Journal of Plasticity}\ }\textbf {\bibinfo {volume} {25}},\
  \bibinfo {pages} {239} (\bibinfo {year} {2009}{\natexlab{a}})}\BibitemShut
  {NoStop}%
\bibitem [{\citenamefont {Levitas}\ and\ \citenamefont
  {Ozsoy}(2009{\natexlab{b}})}]{levitas2009micromechanical2}%
  \BibitemOpen
  \bibfield  {author} {\bibinfo {author} {\bibfnamefont {V.~I.}\ \bibnamefont
  {Levitas}}\ and\ \bibinfo {author} {\bibfnamefont {I.~B.}\ \bibnamefont
  {Ozsoy}},\ }\href@noop {} {\bibfield  {journal} {\bibinfo  {journal}
  {International Journal of Plasticity}\ }\textbf {\bibinfo {volume} {25}},\
  \bibinfo {pages} {546} (\bibinfo {year} {2009}{\natexlab{b}})}\BibitemShut
  {NoStop}%
\bibitem [{\citenamefont {Cayron}(2013)}]{cayron_2013}%
  \BibitemOpen
  \bibfield  {author} {\bibinfo {author} {\bibfnamefont {C.}~\bibnamefont
  {Cayron}},\ }\href@noop {} {\bibfield  {journal} {\bibinfo  {journal} {Acta
  Crystallographica Section A: Foundations of Crystallography}\ }\textbf
  {\bibinfo {volume} {69}},\ \bibinfo {pages} {498} (\bibinfo {year}
  {2013})}\BibitemShut {NoStop}%
\bibitem [{\citenamefont {Cayron}(2015)}]{cayron2015continuous}%
  \BibitemOpen
  \bibfield  {author} {\bibinfo {author} {\bibfnamefont {C.}~\bibnamefont
  {Cayron}},\ }\href@noop {} {\bibfield  {journal} {\bibinfo  {journal} {Acta
  Materialia}\ }\textbf {\bibinfo {volume} {96}},\ \bibinfo {pages} {189}
  (\bibinfo {year} {2015})}\BibitemShut {NoStop}%
\bibitem [{\citenamefont {Baur}\ \emph {et~al.}(2017)\citenamefont {Baur},
  \citenamefont {Cayron},\ and\ \citenamefont {Log{\'e}}}]{baur2017225}%
  \BibitemOpen
  \bibfield  {author} {\bibinfo {author} {\bibfnamefont {A.~P.}\ \bibnamefont
  {Baur}}, \bibinfo {author} {\bibfnamefont {C.}~\bibnamefont {Cayron}}, \ and\
  \bibinfo {author} {\bibfnamefont {R.~E.}\ \bibnamefont {Log{\'e}}},\
  }\href@noop {} {\bibfield  {journal} {\bibinfo  {journal} {Scientific
  reports}\ }\textbf {\bibinfo {volume} {7}},\ \bibinfo {pages} {40938}
  (\bibinfo {year} {2017})}\BibitemShut {NoStop}%
\bibitem [{\citenamefont {Koumatos}\ and\ \citenamefont
  {Muehlemann}(2017)}]{koumatos2017theoretical}%
  \BibitemOpen
  \bibfield  {author} {\bibinfo {author} {\bibfnamefont {K.}~\bibnamefont
  {Koumatos}}\ and\ \bibinfo {author} {\bibfnamefont {A.}~\bibnamefont
  {Muehlemann}},\ }\href@noop {} {\bibfield  {journal} {\bibinfo  {journal}
  {Acta Crystallographica Section A: Foundations and Advances}\ }\textbf
  {\bibinfo {volume} {73}},\ \bibinfo {pages} {115} (\bibinfo {year}
  {2017})}\BibitemShut {NoStop}%
\bibitem [{\citenamefont {Koumatos}\ and\ \citenamefont
  {Muehlemann}(2016)}]{koumatos2016optimality}%
  \BibitemOpen
  \bibfield  {author} {\bibinfo {author} {\bibfnamefont {K.}~\bibnamefont
  {Koumatos}}\ and\ \bibinfo {author} {\bibfnamefont {A.}~\bibnamefont
  {Muehlemann}},\ }\href@noop {} {\bibfield  {journal} {\bibinfo  {journal}
  {Proceedings of the Royal Society A: Mathematical, Physical and Engineering
  Sciences}\ }\textbf {\bibinfo {volume} {472}},\ \bibinfo {pages} {20150865}
  (\bibinfo {year} {2016})}\BibitemShut {NoStop}%
\bibitem [{\citenamefont {Koumatos}\ and\ \citenamefont
  {Muehlemann}(2019)}]{koumatos2019parameter}%
  \BibitemOpen
  \bibfield  {author} {\bibinfo {author} {\bibfnamefont {K.}~\bibnamefont
  {Koumatos}}\ and\ \bibinfo {author} {\bibfnamefont {A.}~\bibnamefont
  {Muehlemann}},\ }\href@noop {} {\bibfield  {journal} {\bibinfo  {journal}
  {Acta Crystallographica Section A: Foundations and Advances}\ }\textbf
  {\bibinfo {volume} {75}} (\bibinfo {year} {2019})}\BibitemShut {NoStop}%
\bibitem [{\citenamefont {Ou}(2017)}]{ou2017molecular}%
  \BibitemOpen
  \bibfield  {author} {\bibinfo {author} {\bibfnamefont {X.}~\bibnamefont
  {Ou}},\ }\href@noop {} {\bibfield  {journal} {\bibinfo  {journal} {Materials
  Science and Technology}\ }\textbf {\bibinfo {volume} {33}},\ \bibinfo {pages}
  {822} (\bibinfo {year} {2017})}\BibitemShut {NoStop}%
\bibitem [{\citenamefont {Stevanovi\'c}\ \emph {et~al.}(2018)\citenamefont
  {Stevanovi\'c}, \citenamefont {Trottier}, \citenamefont {Musgrave},
  \citenamefont {Therrien}, \citenamefont {Holder},\ and\ \citenamefont
  {Graf}}]{stevanovic2018}%
  \BibitemOpen
  \bibfield  {author} {\bibinfo {author} {\bibfnamefont {V.}~\bibnamefont
  {Stevanovi\'c}}, \bibinfo {author} {\bibfnamefont {R.}~\bibnamefont
  {Trottier}}, \bibinfo {author} {\bibfnamefont {C.}~\bibnamefont {Musgrave}},
  \bibinfo {author} {\bibfnamefont {F.}~\bibnamefont {Therrien}}, \bibinfo
  {author} {\bibfnamefont {A.}~\bibnamefont {Holder}}, \ and\ \bibinfo {author}
  {\bibfnamefont {P.}~\bibnamefont {Graf}},\ }\href@noop {} {\bibfield
  {journal} {\bibinfo  {journal} {Phys. Rev. Materials}\ }\textbf {\bibinfo
  {volume} {2}} (\bibinfo {year} {2018})}\BibitemShut {NoStop}%
\bibitem [{\citenamefont {Therrien}\ \emph {et~al.}(2020)\citenamefont
  {Therrien}, \citenamefont {Graf},\ and\ \citenamefont
  {Stevanović}}]{therrien2019}%
  \BibitemOpen
  \bibfield  {author} {\bibinfo {author} {\bibfnamefont {F.}~\bibnamefont
  {Therrien}}, \bibinfo {author} {\bibfnamefont {P.}~\bibnamefont {Graf}}, \
  and\ \bibinfo {author} {\bibfnamefont {V.}~\bibnamefont {Stevanović}},\
  }\href@noop {} {\bibfield  {journal} {\bibinfo  {journal} {The Journal of
  Chemical Physics}\ }\textbf {\bibinfo {volume} {152}},\ \bibinfo {pages}
  {074106} (\bibinfo {year} {2020})}\BibitemShut {NoStop}%
\bibitem [{Note1()}]{Note1}%
  \BibitemOpen
  \bibinfo {note} {Github.com/ftherrien/p2ptrans}\BibitemShut {NoStop}%
\bibitem [{\citenamefont {Bogers}\ and\ \citenamefont
  {Burgers}(1964)}]{bogers1964partial}%
  \BibitemOpen
  \bibfield  {author} {\bibinfo {author} {\bibfnamefont {A.}~\bibnamefont
  {Bogers}}\ and\ \bibinfo {author} {\bibfnamefont {W.}~\bibnamefont
  {Burgers}},\ }\href@noop {} {\bibfield  {journal} {\bibinfo  {journal} {Acta
  Metallurgica}\ }\textbf {\bibinfo {volume} {12}},\ \bibinfo {pages} {255}
  (\bibinfo {year} {1964})}\BibitemShut {NoStop}%
\bibitem [{\citenamefont {Ong}\ \emph {et~al.}(2013)\citenamefont {Ong},
  \citenamefont {Richards}, \citenamefont {Jain}, \citenamefont {Hautier},
  \citenamefont {Kocher}, \citenamefont {Cholia}, \citenamefont {Gunter},
  \citenamefont {Chevrier}, \citenamefont {Persson},\ and\ \citenamefont
  {Ceder}}]{ong2013python}%
  \BibitemOpen
  \bibfield  {author} {\bibinfo {author} {\bibfnamefont {S.~P.}\ \bibnamefont
  {Ong}}, \bibinfo {author} {\bibfnamefont {W.~D.}\ \bibnamefont {Richards}},
  \bibinfo {author} {\bibfnamefont {A.}~\bibnamefont {Jain}}, \bibinfo {author}
  {\bibfnamefont {G.}~\bibnamefont {Hautier}}, \bibinfo {author} {\bibfnamefont
  {M.}~\bibnamefont {Kocher}}, \bibinfo {author} {\bibfnamefont
  {S.}~\bibnamefont {Cholia}}, \bibinfo {author} {\bibfnamefont
  {D.}~\bibnamefont {Gunter}}, \bibinfo {author} {\bibfnamefont {V.~L.}\
  \bibnamefont {Chevrier}}, \bibinfo {author} {\bibfnamefont {K.~A.}\
  \bibnamefont {Persson}}, \ and\ \bibinfo {author} {\bibfnamefont
  {G.}~\bibnamefont {Ceder}},\ }\href@noop {} {\bibfield  {journal} {\bibinfo
  {journal} {Computational Materials Science}\ }\textbf {\bibinfo {volume}
  {68}},\ \bibinfo {pages} {314} (\bibinfo {year} {2013})}\BibitemShut
  {NoStop}%
\bibitem [{\citenamefont {Hall}(2015)}]{hall2015matrix}%
  \BibitemOpen
  \bibfield  {author} {\bibinfo {author} {\bibfnamefont {B.~C.}\ \bibnamefont
  {Hall}},\ }in\ \href@noop {} {\emph {\bibinfo {booktitle} {Lie Groups, Lie
  Algebras, and Representations}}}\ (\bibinfo  {publisher} {Springer},\
  \bibinfo {year} {2015})\ pp.\ \bibinfo {pages} {31--48}\BibitemShut {NoStop}%
\bibitem [{\citenamefont {Ro{\u\i}tburd}(1993)}]{roitburd1993}%
  \BibitemOpen
  \bibfield  {author} {\bibinfo {author} {\bibfnamefont {A.~L.}\ \bibnamefont
  {Ro{\u\i}tburd}},\ }\href@noop {} {\bibfield  {journal} {\bibinfo  {journal}
  {Phase Transitions}\ }\textbf {\bibinfo {volume} {45}},\ \bibinfo {pages} {1}
  (\bibinfo {year} {1993})}\BibitemShut {NoStop}%
\bibitem [{\citenamefont {Kelly}(1953)}]{kelly1953neumann}%
  \BibitemOpen
  \bibfield  {author} {\bibinfo {author} {\bibfnamefont {A.}~\bibnamefont
  {Kelly}},\ }\href@noop {} {\bibfield  {journal} {\bibinfo  {journal}
  {Proceedings of the Physical Society. Section A}\ }\textbf {\bibinfo {volume}
  {66}},\ \bibinfo {pages} {403} (\bibinfo {year} {1953})}\BibitemShut
  {NoStop}%
\bibitem [{\citenamefont {Shimizu}\ \emph {et~al.}(1970)\citenamefont
  {Shimizu}, \citenamefont {Oka},\ and\ \citenamefont
  {Wayman}}]{shimizu1970association}%
  \BibitemOpen
  \bibfield  {author} {\bibinfo {author} {\bibfnamefont {K.}~\bibnamefont
  {Shimizu}}, \bibinfo {author} {\bibfnamefont {M.}~\bibnamefont {Oka}}, \ and\
  \bibinfo {author} {\bibfnamefont {C.}~\bibnamefont {Wayman}},\ }\href@noop {}
  {\bibfield  {journal} {\bibinfo  {journal} {Acta Metallurgica}\ }\textbf
  {\bibinfo {volume} {18}},\ \bibinfo {pages} {1005} (\bibinfo {year}
  {1970})}\BibitemShut {NoStop}%
\bibitem [{\citenamefont {Zhang}\ and\ \citenamefont
  {Kelly}(2005)}]{zhang_kelly_2005}%
  \BibitemOpen
  \bibfield  {author} {\bibinfo {author} {\bibfnamefont {M.-X.}\ \bibnamefont
  {Zhang}}\ and\ \bibinfo {author} {\bibfnamefont {P.}~\bibnamefont {Kelly}},\
  }\href {\doibase https://doi.org/10.1016/j.scriptamat.2005.01.040} {\bibfield
   {journal} {\bibinfo  {journal} {Scripta Materialia}\ }\textbf {\bibinfo
  {volume} {52}},\ \bibinfo {pages} {963 } (\bibinfo {year}
  {2005})}\BibitemShut {NoStop}%
\bibitem [{\citenamefont {Bl{\"o}chl}(1994)}]{blochl1994projector}%
  \BibitemOpen
  \bibfield  {author} {\bibinfo {author} {\bibfnamefont {P.~E.}\ \bibnamefont
  {Bl{\"o}chl}},\ }\href@noop {} {\bibfield  {journal} {\bibinfo  {journal}
  {Physical review B}\ }\textbf {\bibinfo {volume} {50}},\ \bibinfo {pages}
  {17953} (\bibinfo {year} {1994})}\BibitemShut {NoStop}%
\bibitem [{\citenamefont {Kresse}\ and\ \citenamefont
  {Furthm{\"u}ller}(1996)}]{kresse1996efficiency}%
  \BibitemOpen
  \bibfield  {author} {\bibinfo {author} {\bibfnamefont {G.}~\bibnamefont
  {Kresse}}\ and\ \bibinfo {author} {\bibfnamefont {J.}~\bibnamefont
  {Furthm{\"u}ller}},\ }\href@noop {} {\bibfield  {journal} {\bibinfo
  {journal} {Computational materials science}\ }\textbf {\bibinfo {volume}
  {6}},\ \bibinfo {pages} {15} (\bibinfo {year} {1996})}\BibitemShut {NoStop}%
\bibitem [{\citenamefont {Bollmann}(1970)}]{bollmann1}%
  \BibitemOpen
  \bibfield  {author} {\bibinfo {author} {\bibfnamefont {W.}~\bibnamefont
  {Bollmann}},\ }\href@noop {} {\emph {\bibinfo {title} {Crystal Defects and
  Crystalline Interfaces}}},\ \bibinfo {edition} {1st}\ ed.\ (\bibinfo
  {publisher} {Springer-Verlag Berlin Heidelberg},\ \bibinfo {year} {1970})\
  pp.\ \bibinfo {pages} {78--97}\BibitemShut {NoStop}%
\bibitem [{\citenamefont {Balluffi}\ \emph {et~al.}(1982)\citenamefont
  {Balluffi}, \citenamefont {Brokman},\ and\ \citenamefont
  {King}}]{balluffi_brokman_king_1982}%
  \BibitemOpen
  \bibfield  {author} {\bibinfo {author} {\bibfnamefont {R.}~\bibnamefont
  {Balluffi}}, \bibinfo {author} {\bibfnamefont {A.}~\bibnamefont {Brokman}}, \
  and\ \bibinfo {author} {\bibfnamefont {A.}~\bibnamefont {King}},\ }\href
  {\doibase 10.1016/0001-6160(82)90166-3} {\bibfield  {journal} {\bibinfo
  {journal} {Acta Metallurgica}\ }\textbf {\bibinfo {volume} {30}},\ \bibinfo
  {pages} {1453} (\bibinfo {year} {1982})}\BibitemShut {NoStop}%
\bibitem [{\citenamefont {Zur}\ and\ \citenamefont {McGill}(1984)}]{zur1984}%
  \BibitemOpen
  \bibfield  {author} {\bibinfo {author} {\bibfnamefont {A.}~\bibnamefont
  {Zur}}\ and\ \bibinfo {author} {\bibfnamefont {T.~C.}\ \bibnamefont
  {McGill}},\ }\href {\doibase 10.1063/1.333084} {\bibfield  {journal}
  {\bibinfo  {journal} {Journal of Applied Physics}\ }\textbf {\bibinfo
  {volume} {55}},\ \bibinfo {pages} {378} (\bibinfo {year} {1984})}\BibitemShut
  {NoStop}%
\bibitem [{\citenamefont {Ikuhara}\ and\ \citenamefont
  {Pirouz}(1996)}]{ikuhara_pirouz_1996}%
  \BibitemOpen
  \bibfield  {author} {\bibinfo {author} {\bibfnamefont {Y.}~\bibnamefont
  {Ikuhara}}\ and\ \bibinfo {author} {\bibfnamefont {P.}~\bibnamefont
  {Pirouz}},\ }\href {\doibase 10.4028/www.scientific.net/msf.207-209.121}
  {\bibfield  {journal} {\bibinfo  {journal} {Materials Science Forum}\
  }\textbf {\bibinfo {volume} {207-209}},\ \bibinfo {pages} {121} (\bibinfo
  {year} {1996})}\BibitemShut {NoStop}%
\bibitem [{\citenamefont {Tkatchenko}\ and\ \citenamefont
  {Batina}(2004)}]{tkatchenko2004unequal}%
  \BibitemOpen
  \bibfield  {author} {\bibinfo {author} {\bibfnamefont {A.}~\bibnamefont
  {Tkatchenko}}\ and\ \bibinfo {author} {\bibfnamefont {N.}~\bibnamefont
  {Batina}},\ }\href@noop {} {\bibfield  {journal} {\bibinfo  {journal}
  {Physical Review B}\ }\textbf {\bibinfo {volume} {70}},\ \bibinfo {pages}
  {195403} (\bibinfo {year} {2004})}\BibitemShut {NoStop}%
\bibitem [{\citenamefont {Tkatchenko}\ and\ \citenamefont
  {Batina}(2005)}]{tkatchenko2005unequal}%
  \BibitemOpen
  \bibfield  {author} {\bibinfo {author} {\bibfnamefont {A.}~\bibnamefont
  {Tkatchenko}}\ and\ \bibinfo {author} {\bibfnamefont {N.}~\bibnamefont
  {Batina}},\ }\href@noop {} {\bibfield  {journal} {\bibinfo  {journal} {The
  Journal of Physical Chemistry B}\ }\textbf {\bibinfo {volume} {109}},\
  \bibinfo {pages} {21710} (\bibinfo {year} {2005})}\BibitemShut {NoStop}%
\bibitem [{\citenamefont {Tkatchenko}\ and\ \citenamefont
  {Batina}(2006)}]{tkatchenko2006classification}%
  \BibitemOpen
  \bibfield  {author} {\bibinfo {author} {\bibfnamefont {A.}~\bibnamefont
  {Tkatchenko}}\ and\ \bibinfo {author} {\bibfnamefont {N.}~\bibnamefont
  {Batina}},\ }\href@noop {} {\bibfield  {journal} {\bibinfo  {journal} {The
  Journal of chemical physics}\ }\textbf {\bibinfo {volume} {125}},\ \bibinfo
  {pages} {164702} (\bibinfo {year} {2006})}\BibitemShut {NoStop}%
\bibitem [{\citenamefont {Ding}\ \emph {et~al.}(2016)\citenamefont {Ding},
  \citenamefont {Dwaraknath}, \citenamefont {Garten}, \citenamefont {Ndione},
  \citenamefont {Ginley},\ and\ \citenamefont {Persson}}]{persson_ACS:2016}%
  \BibitemOpen
  \bibfield  {author} {\bibinfo {author} {\bibfnamefont {H.}~\bibnamefont
  {Ding}}, \bibinfo {author} {\bibfnamefont {S.~S.}\ \bibnamefont
  {Dwaraknath}}, \bibinfo {author} {\bibfnamefont {L.}~\bibnamefont {Garten}},
  \bibinfo {author} {\bibfnamefont {P.}~\bibnamefont {Ndione}}, \bibinfo
  {author} {\bibfnamefont {D.}~\bibnamefont {Ginley}}, \ and\ \bibinfo {author}
  {\bibfnamefont {K.~A.}\ \bibnamefont {Persson}},\ }\bibfield  {booktitle}
  {\emph {\bibinfo {booktitle} {ACS Applied Materials \& Interfaces}},\ }\href
  {\doibase 10.1021/acsami.6b01630} {\bibfield  {journal} {\bibinfo  {journal}
  {ACS Applied Materials \& Interfaces}\ }\textbf {\bibinfo {volume} {8}},\
  \bibinfo {pages} {13086} (\bibinfo {year} {2016})}\BibitemShut {NoStop}%
\bibitem [{\citenamefont {Jelver}\ \emph {et~al.}(2017)\citenamefont {Jelver},
  \citenamefont {Larsen}, \citenamefont {Stradi}, \citenamefont {Stokbro},\
  and\ \citenamefont {Jacobsen}}]{jelver2017determination}%
  \BibitemOpen
  \bibfield  {author} {\bibinfo {author} {\bibfnamefont {L.}~\bibnamefont
  {Jelver}}, \bibinfo {author} {\bibfnamefont {P.~M.}\ \bibnamefont {Larsen}},
  \bibinfo {author} {\bibfnamefont {D.}~\bibnamefont {Stradi}}, \bibinfo
  {author} {\bibfnamefont {K.}~\bibnamefont {Stokbro}}, \ and\ \bibinfo
  {author} {\bibfnamefont {K.~W.}\ \bibnamefont {Jacobsen}},\ }\href@noop {}
  {\bibfield  {journal} {\bibinfo  {journal} {Physical Review B}\ }\textbf
  {\bibinfo {volume} {96}},\ \bibinfo {pages} {085306} (\bibinfo {year}
  {2017})}\BibitemShut {NoStop}%
\end{thebibliography}

\end{document}